\documentclass[twocolumn,showpacs,aps,prl,amsmath,amssymb,nobibnotes,floatfix]{revtex4-1}
\newcommand{\beq}{\begin{equation}}
\newcommand{\eeq}{\end{equation}}
\newcommand{\beqn}{\begin{eqnarray}}
\newcommand{\eeqn}{\end{eqnarray}}

\newcommand{\apjl}{Astrophys.\ J.\ Lett.}
\newcommand{\apjs}{Astrophys.\ J.\ Supp.}
\newcommand{\mnras}{Mon. Not. Roy. Astron. Soc.}

\newcommand{\llabel}[1]{\label{#1}}              
\newcommand{\labeq}[2]{ \begin{equation} \llabel{#1}
{#2}
\end{equation}}
\usepackage{graphicx}
\usepackage{epsf}
\usepackage{graphics,epsfig,placeins,subfigure,wrapfig}
\begin{document}
\title{General relativistic simulations of binary black hole-neutron stars:\\ Precursor electromagnetic signals}

\author{Vasileios Paschalidis}
\author{Zachariah B. Etienne}
\author{Stuart~L.~Shapiro}
\altaffiliation{Also at Department of Astronomy and NCSA, University of
  Illinois at Urbana-Champaign, Urbana, IL 61801}
\affiliation{Department of Physics, University of Illinois at
  Urbana-Champaign, Urbana, IL 61801}

\begin{abstract}

We perform the first general relativistic force-free simulations of
neutron star magnetospheres in orbit about spinning and non-spinning
black holes.  We find promising precursor electromagnetic emission:
typical Poynting luminosities at, e.g., an orbital separation
of $r=6.6R_{\rm NS}$ are $L_{\rm EM} \sim 6\times 10^{42}(B_{\rm
  NS,p}/10^{13}\rm G)^2(M_{\rm NS}/1.4M_\odot)^2$erg/s. The Poynting
flux peaks within a broad beam of $\sim 40^\circ$ in the azimuthal
direction and within $\sim 60^\circ$ from the orbital plane,
establishing a possible lighthouse effect. Our calculations, though
preliminary, preview more detailed simulations of these systems that
we plan to perform in the future.

\end{abstract}


\maketitle

Black hole--neutron star (BHNS) binaries are promising
sources for the simultaneous detection of gravitational wave (GW) and
electromagnetic (EM) signals in the era of multimessenger
astronomy. For example, aLIGO is expected to detect between
1--100 BHNS GW signals each
year~\cite{LIGO1,LIGO2,KBKOW,aetal10}. Furthermore, BHNS mergers may
provide the central engine powering a short-hard gamma-ray burst
(sGRB). GW signals from the inspiral and merger of BHNSs were
computed recently in full general relativity (GR)
\cite{UIUC_BHNS__BH_SPIN_PAPER,Duez:2009yy,Foucart:2010eq,Kyutoku:2011vz,Lackey:2011vz,Foucart:2012vn,Lovelace:2013vma},
and the first parametric study of magnetized BHNS mergers in full GR
has been carried out in
\cite{UIUC_MAGNETIZED_BHNS_PAPER1,UIUC_MAGNETIZED_BHNS_PAPER2}, where
it was shown that under appropriate conditions BHNSs can launch
collimated jets -- necessary ingredients for many sGRB models.

Detecting pre-merger EM signals, combined with GW observations, will
yield a wealth of information about BHNS binaries. EM signals will
help localize the source on the sky, resulting in improved parameter
estimation from GWs
~\cite{Nissanke:2012dj}.

Neutron stars likely possess dipole magnetic fields and a force-free
magnetosphere \cite{GJ1969}. Toward the end of a BHNS inspiral, strong
magnetic fields will sweep the BH, possibly establishing a unipolar
inductor (UI) that extracts energy from the system
\cite{UI1969,Drell1965}. This exciting new possibility has been
suggested recently as a potential mechanism for powering precursor EM
signals from BHNSs \cite{McL2011}. Follow-up analytical approximations
in the high-mass-ratio limit have been performed
\cite{Lyutikov:2011tq,OL2013} to estimate the output power. But, as
these UIs operate in strongly-curved, dynamical spacetimes, numerical
relativity simulations are necessary to reliably determine the amount
of EM output, particularly in the regime of comparable-mass binaries
where previous approximations do not apply. While UIs may also exist
in NSNS binaries
\cite{Hansen:2000am,Lyutikov:2011tq,2012ApJ...755...80P,Palenzuela:2013hu},
BHNSs may be optimal systems for this mechanism because the azimuthal
twist ($\zeta_\phi$) of the magnetic flux tubes is less than unity for
a BH resistor \cite{DL2012}.


In this paper we simulate NS magnetospheres in orbit about spinning
and nonspinning BHs prior to merger via general relativistic,
force-free (GRFF) simulations. We calculate the Poynting luminosity
and characterize its angular dependence. We also treat another
EM emission mechanism: magnetic dipole (MD) radiation from the
accelerating NS. MD radiation has been considered in the context of
EM emission affecting the inspiral and GW signal \cite{Ioka:2000yb},
but not as a source for strong precursor EM signals. Here we
show that the MD Poynting luminosity is significant, and may
dominate the EM output in cases where UI ceases
 due to corotation or $\zeta_\phi>1$.
We use geometrized units where $c=1=G$, unless
otherwise stated.


\begin{figure*}
\centering
\includegraphics[width=0.33\textwidth]{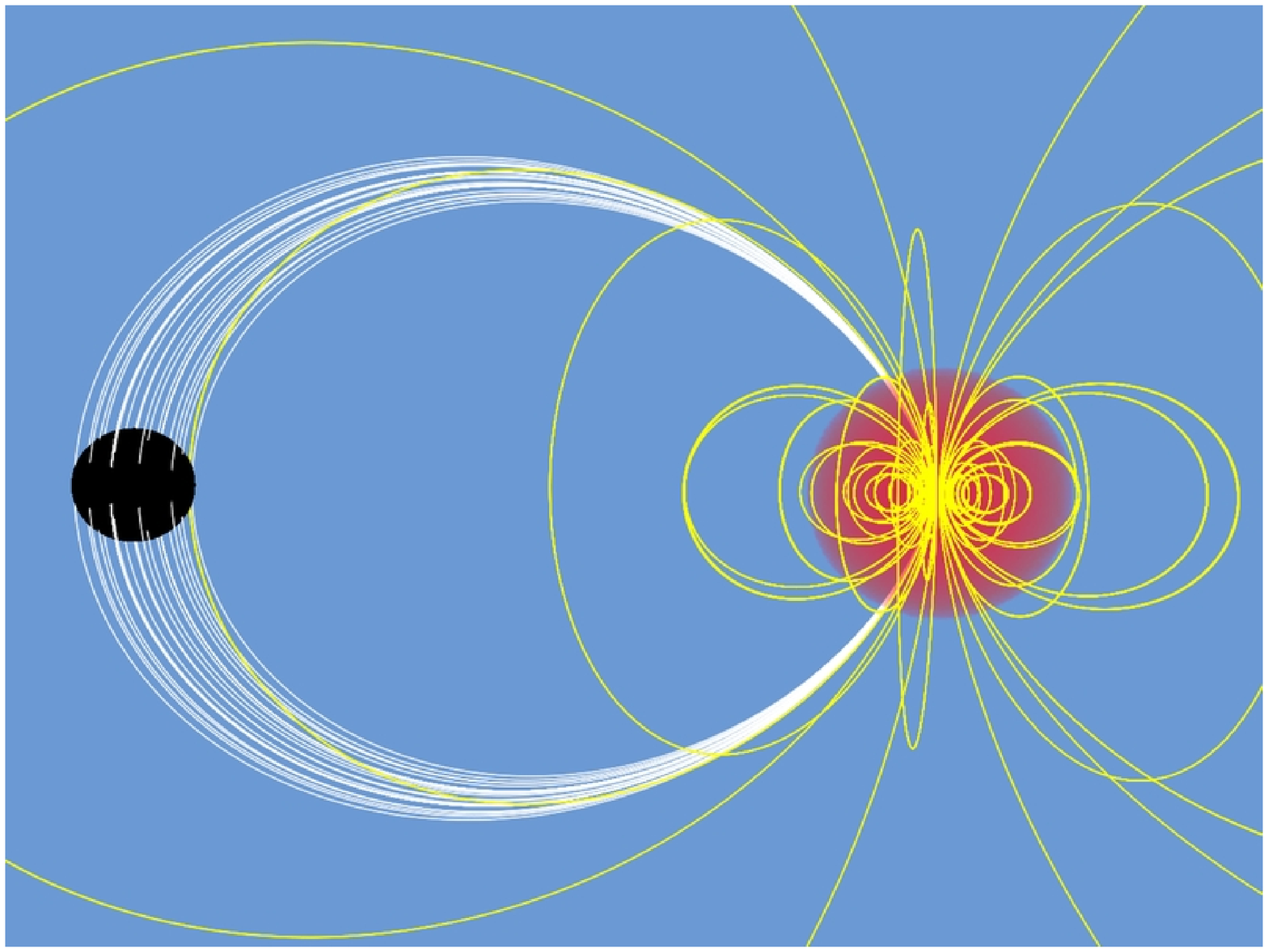}
\includegraphics[width=0.33\textwidth]{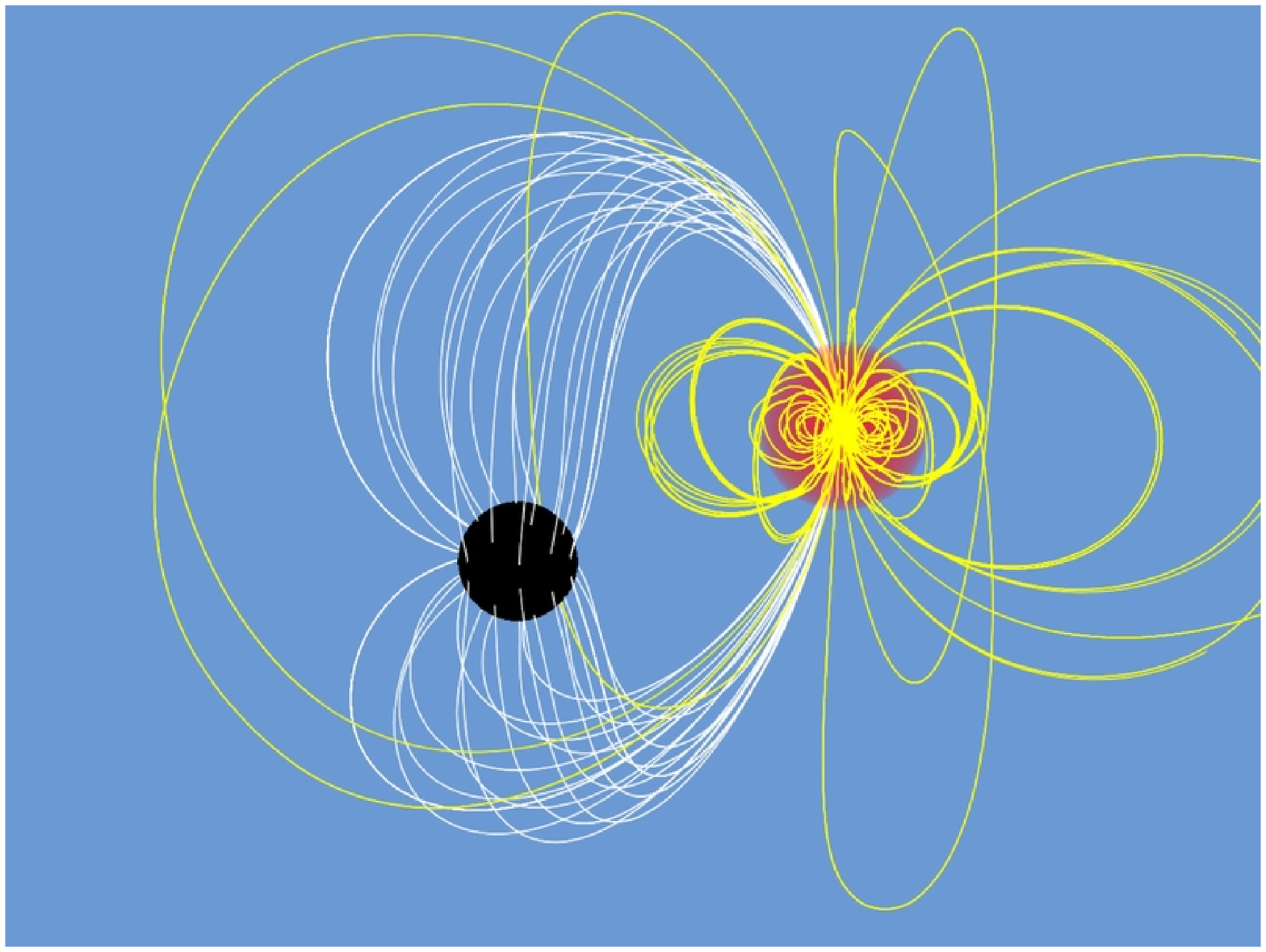}
\includegraphics[width=0.33\textwidth]{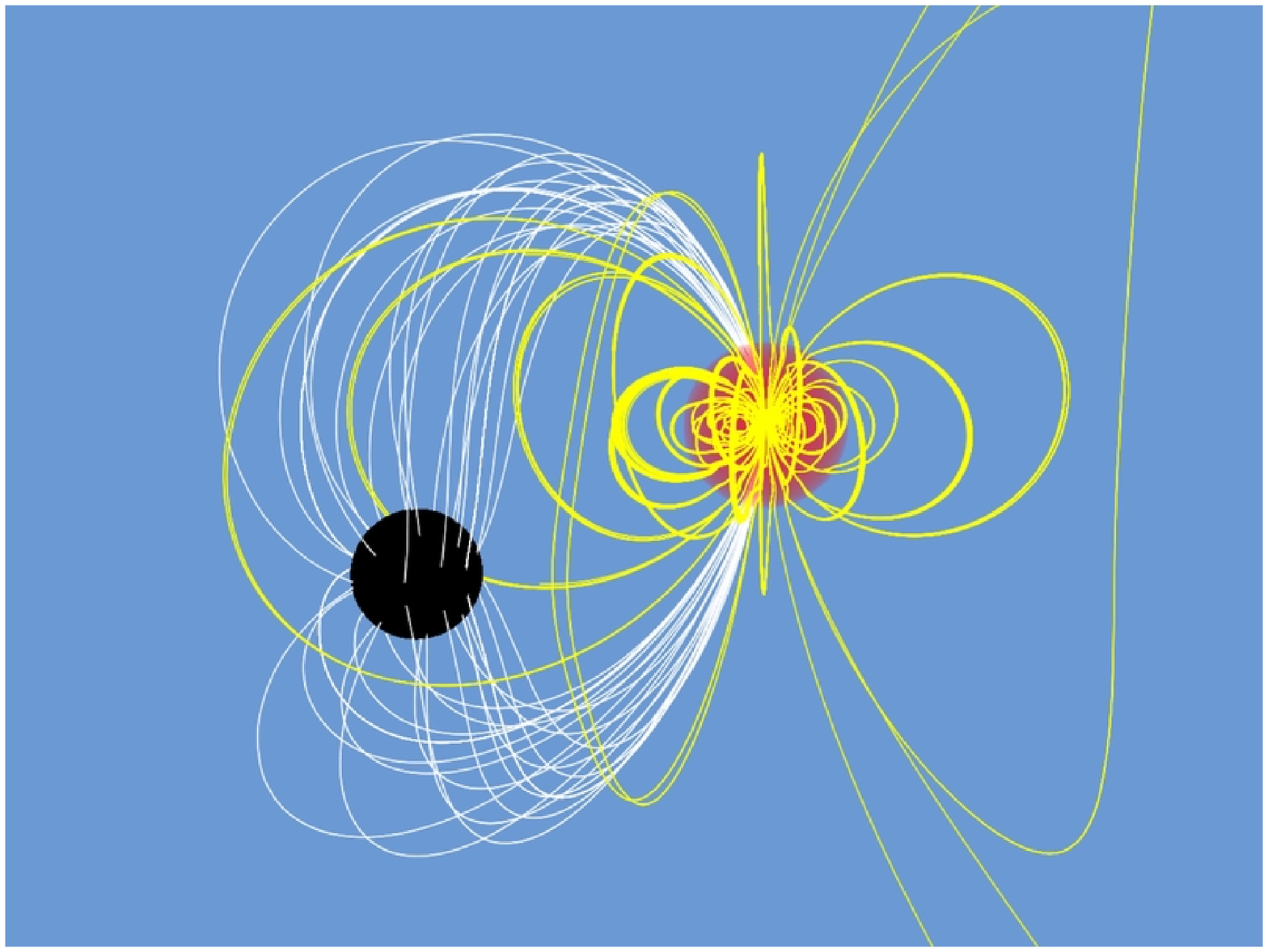}
\includegraphics[width=0.33\textwidth]{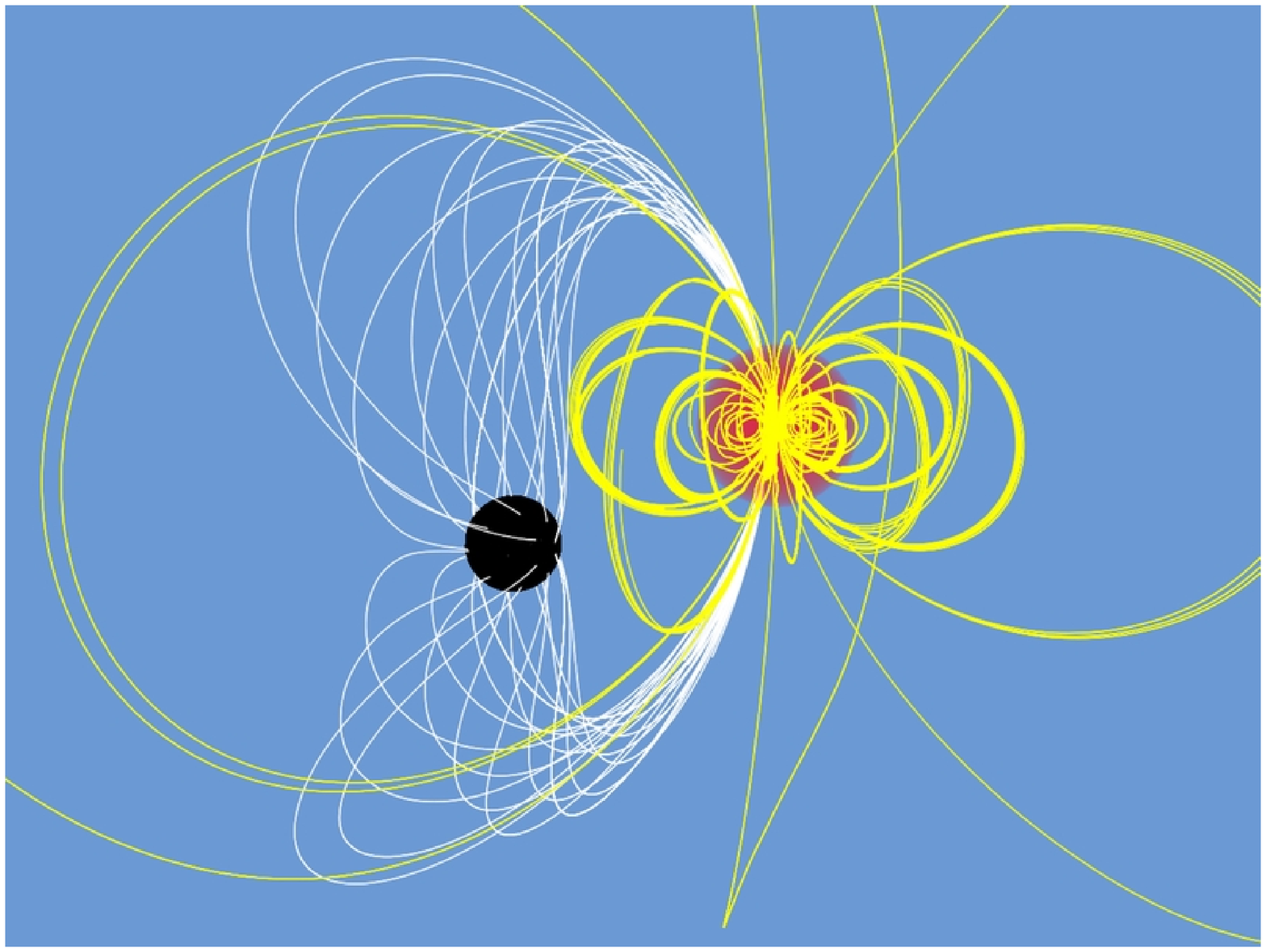}
\caption{Initial magnetic field in the $a_\ast=0$ case (upper
  left panel). Relaxed magnetic field at $t\approx1.5$ orbits: $a_\ast=0$
  (lower left panel), $a_\ast=-0.5$ (upper right panel), and $a_\ast=0.75$
  (lower right panel). The black sphere represents the BH horizon and
  the NS is shown in red. Both white and yellow lines are the magnetic
  fields lines. White lines distinguish field lines that intersect
  the BH horizon.}
\label{Fig:Binitial_and_final}
\end{figure*}


Solving the GRFF equations generally involves evolving the electric
(${\bf E}$) and magnetic (${\bf B}$) fields under the force-free
constraints ${\bf E}\cdot {\bf B}=0$ and $E^2 < B^2$
\cite{Komissarov:2002my,2010PhRvD..82d4045P}. The force-free regime
represents the limit of ideal MHD when the magnetic fields dominate
the plasma dynamics \cite{Komissarov:2002my,McKinney:2004ka}. In this
regime, one can choose the ${\bf B}$-field and the Poynting vector
${\bf S}$ as dynamical variables, and cast their evolution
equations in conservation form \cite{2006MNRAS.367.1797M,UIUC_GRFFE}.
The force-free constraints then become ${\bf S}\cdot {\bf B} = 0$ and
$S^2 < B^4$ 
\cite{UIUC_GRFFE}. An advantage of this formulation is that it can be
easily embedded into an ideal GRMHD code
\cite{2006MNRAS.367.1797M}. The GRFF formulation adopted here 
is identical to \cite{2006MNRAS.367.1797M}, except that at every
timestep, in addition to $S^2 < B^4$, we also enforce the algebraic
constraint ${\bf S}\cdot {\bf B} = 0$, which was ignored in
\cite{2006MNRAS.367.1797M}. For a discussion of possible shortcomings
of this corrective enforcement of the force-free conditions see
\cite{2006MNRAS.367.1797M,Spitkovsky:2006np}.  This formulation is
embedded in the fully GRMHD infrastructure presented and tested in
\cite{Duez:2005sf,Etienne:2010ui,UIUCEMGAUGEPAPER}.  Moreover, to
enforce the ${\bf \nabla}\cdot {\bf B} =0$ constraint on our
adaptive-mesh-refinement grids, the magnetic induction equation is
evolved via the vector potential formulation introduced in
\cite{Etienne:2010ui,UIUCEMGAUGEPAPER,UIUC_MAGNETIZED_BHNS_PAPER1},
coupled to the Generalized Lorenz (GL) gauge condition
\cite{UIUCEMGAUGEPAPER,Farris:2012ux,UIUC_MAGNETIZED_BHNS_PAPER2},
%
with damping parameter $\xi = 1.5/\Delta t$, where $\Delta t$ is the
coarsest level's timestep.

At large separations, the inspiral timescale is much longer than the
orbital timescale.  So to model the BHNS spacetime and the NS matter
fields, we adopt quasiequilibrium solutions of the
conformal-thin-sandwich (CTS) equations for companions at fixed
orbital separation
\cite{Taniguchi:2005fr,UIUC_BHNS__BH_SPIN_PAPER,BSBook}.  The CTS
approximation is excellent at the separations and BH spins considered
here, yielding a binary spacetime with a helical Killing vector. In
such a spacetime the matter and gravitational fields are stationary in
the corotating frame of the binary, enabling us to perform the
simulations in the center-of-mass frame by simply rotating the metric,
as well as the fluid rest-mass density and four-velocity, following
\cite{Farris:2011vx}. This reduces the problem to evolving the EM
fields (${\bf B}$ and ${\bf S}$) in the background matter fields and
spacetime.


Given that force-free electrodynamics is a limit of ideal MHD, the
{\it same} ideal MHD evolution equations can be used to evolve both
the NS interior and the force-free exterior EM fields, provided in the
exterior a compatible force-free velocity is used
\cite{2006MNRAS.367.1797M} and the rest-mass density is set to
zero. This guarantees a smooth transition from the ideal MHD interior
to the force-free exterior, and the MHD variables on the NS surface
effectively provide boundary conditions for the exterior force-free
evolution.
However, given that the chosen initial A-field
is not a CTS solution, we evolve the induction equation [Eqs. (8), (9)
in \cite{UIUC_MAGNETIZED_BHNS_PAPER1}] in the NS interior, 
using the known CTS fluid four-velocity. This sets the boundary condition 
on the NS surface for the Poynting vector and magnetic field in the 
exterior. 
For more details see \cite{UIUC_GRFFE}. An alternative scheme for
matching the interior ideal MHD to the exterior force-free regime was
introduced in \cite{2011arXiv1112.2622L}.  

After tidal disruption, a GRFF treatment becomes inadequate
and must be replaced by full GRMHD. Furthermore, according to 
\cite{MestelShibata1994,Contopoulos1999,Mizuno2013} the ideal MHD
approximation may break down in the regions near the surface. This
motivates a resistive GRMHD simulation with realistic conductivity,
including cooling. However, here we take the widely adopted approach
of neglecting the magnetic field backreaction onto the NS matter
(e.g. \cite{Gruzinov:2004jc,Spitkovsky:2006np,2011arXiv1112.2622L}),
which likely becomes important in a region in the outer layers of the
NS, and assume ideal MHD throughout. Preliminary resistive MHD studies
of rotating neutron stars in \cite{Palenzuela:2012my}, which include
the effects EM backreaction onto the NS matter, show that the outgoing
EM luminosity is within $20\%$ of the values obtained in
\cite{Spitkovsky:2006np}, which neglect the EM backreaction onto the
matter. Thus, we expect the error of neglecting the EM backreaction to
be of this order magnitude at most.


In addition to our new GRFF evolution techniques, we have also added
two equivalent diagnostics to monitor the outgoing EM
luminosity: (i) the $\phi_2$ Newman-Penrose scalar
\cite{NP1962JMP,Teukolsky:1973ha,2010PhRvD..82d4045P}, and (ii) the
Poynting vector ${\bf S}=({\bf E}\times {\bf B})/4\pi$.  To compute
$\phi_2$ we use the same null tetrad as in \cite{Moesta:2011bn}, and
the outgoing luminosity is~\footnote{Notice the factor of $1/4\pi$,
  which was omitted in \cite{Moesta:2011bn}.  This factor results from
  the choice of null tetrad.  A similar factor of $1/2\pi$ was used in
  the $L_{\rm EM}$ formulae reported in
  \cite{Teukolsky:1973ha,2010PhRvD..82d4045P} where a different null
  tetrad was used.}
\labeq{Lout}{ 
L_{\rm EM} \equiv \lim_{r \rightarrow \infty}
  \frac{1}{4\pi}\int r^2|\phi_2|^2 d\Omega=\lim_{r \rightarrow \infty}
  \int r^2S^{\hat r} d\Omega.  
} 
%

\begin{figure*}
\centering
\subfigure{\includegraphics[width=0.255\textwidth]{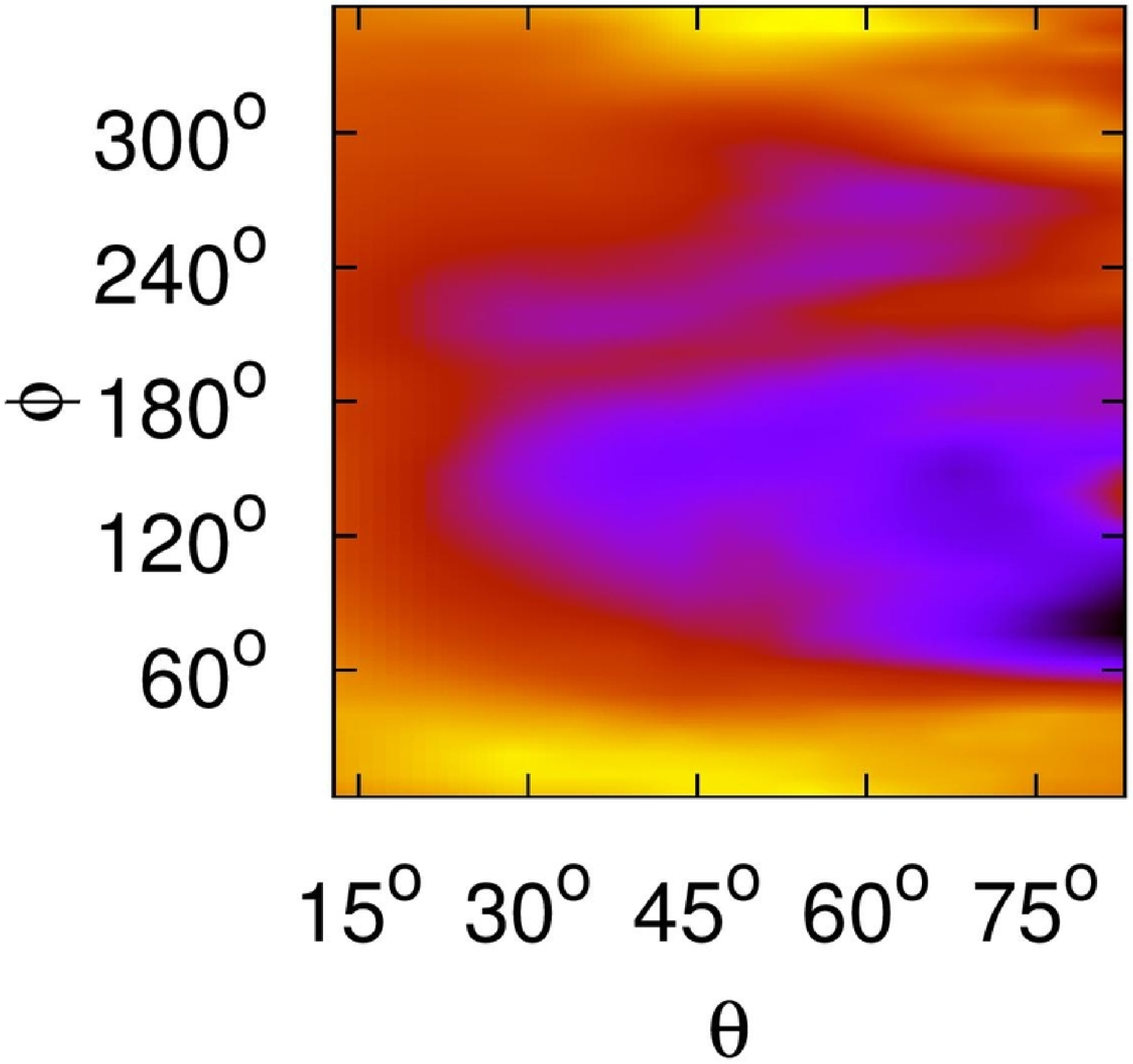}}
\subfigure{\includegraphics[width=0.255\textwidth]{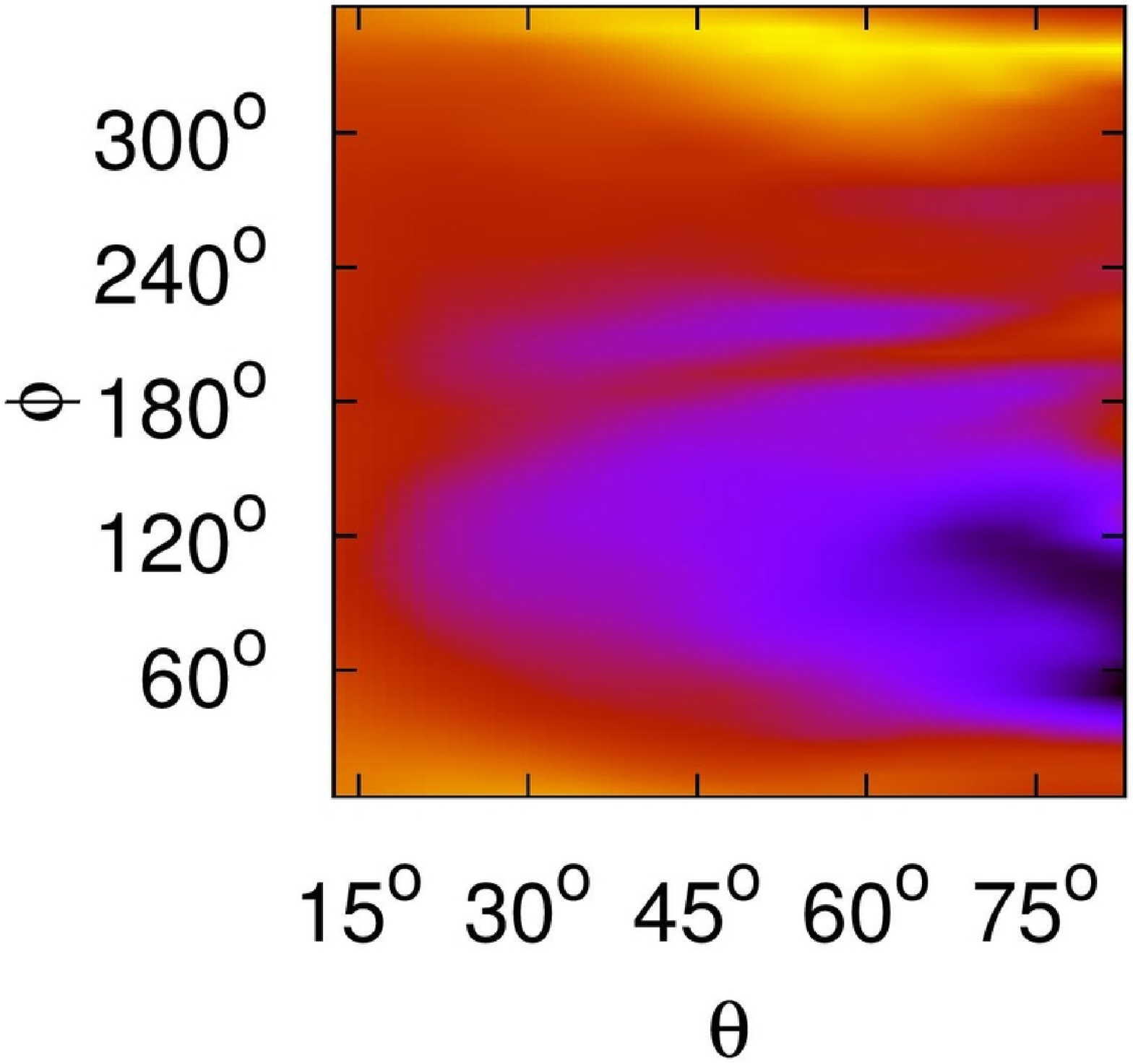}}
\subfigure{\includegraphics[width=0.341\textwidth]{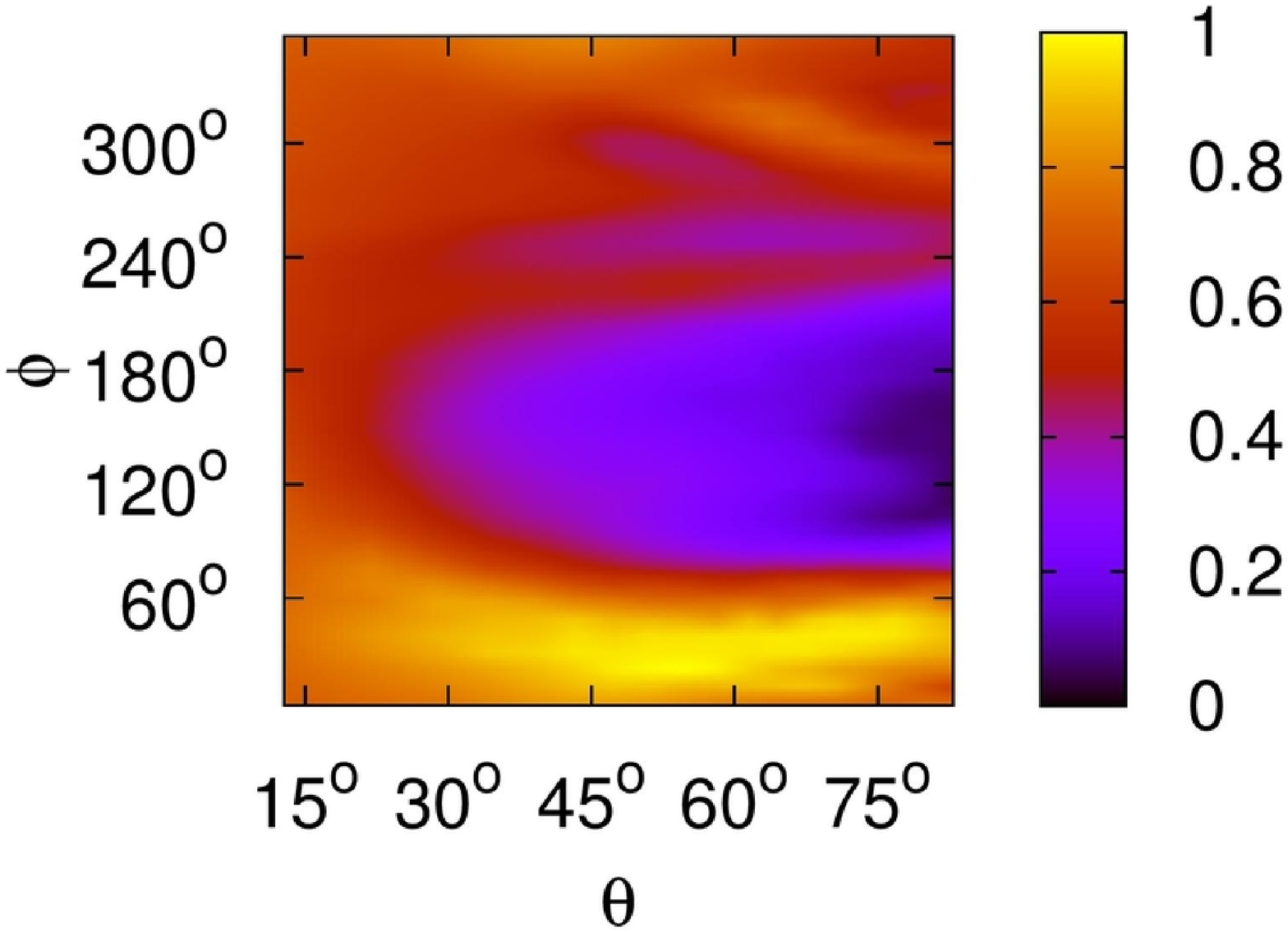}}
\vspace{-0.25cm}
\caption{Angular distribution of Poynting flux, normalized by its peak
  value on a sphere of radius $120M=915(M_{\rm
    NS}/1.4M_\odot)\rm$km. Left: spin -0.5, middle: spin 0, right:
  spin 0.75. The plots correspond to a time after $\sim 2$ orbits. The
  azimuthal ($\phi$) and polar ($\theta$) angles are defined with
  respect to a spherical coordinate system centered on the center of
  mass of the binary.
}
\label{fig:poyntingfluxdistrib}
\centering
\end{figure*}

The spacetime and NS initial data we use correspond to cases A, B, C
in Table I in \cite{UIUC_BHNS__BH_SPIN_PAPER}. The BH spin
parameters are $a_\ast\equiv a/M_{H}=-0.5,0,0.75$, and the BH:NS mass
ratio is $q=3$.
The NS fluid is modeled as an equilibrium, irrotational,
unmagnetized, $\Gamma = 2$ polytrope.  We seed the initial NS with a
purely poloidal magnetic field that approximately corresponds to that
generated by a current loop. The coordinate-basis toroidal component
of this vector potential is
\labeq{Aphi}{
  A_{\phi} = \frac{\pi r_0^2 I_0 \varpi^2}{(r_0^2+r^2)^{3/2}}\bigg(1+\frac{15
    r_0^2(r_0^2+\varpi^2)}{8(r_0^2+r^2)^2}\bigg), 
} 
where $r_0$ is the current loop radius, $I_0$ the loop current,
$r^2=(x-x_{\rm NS})^2+(y-y_{\rm NS})^2+z^2$, $\varpi^2= (x-x_{\rm
  NS})^2+(y-y_{\rm NS})^2$, and $x_{\rm NS}, y_{\rm NS}$ are the
initial coordinates of the NS center of mass. For $r_0 \ll r$
Eq. \eqref{Aphi} gives rise to the standard B-field from a current
loop on the z-axis, and the characteristic $1/r^3$ fall-off of a
standard magnetic dipole on the $z=0$ plane. Choosing $r_0= R_{\rm
  NS}/3$ in all our simulations, where $R_{\rm NS}$ is the NS polar
radius, we find that the initial magnetic field scales as $1/r^3$
outside the NS to a very good degree.  Our simulations scale with $|B|$.
If we set $I_0 = 0.0007$, the initial
NS polar magnetic field (as measured by a CTS normal observer) is
$8.8\times 10^{15}$G. The initial B-field geometry is shown in
the upper left panel of Fig.~\ref{Fig:Binitial_and_final}. To set the
initial electric field, we first set the matter velocity $u_i$ in the
interior according to the CTS solution, and set the exterior $u_i$ to 0
except for the perpendicular component to the B-field, which falls-off
as $1/r^2$ from its NS surface value. The E-field is then
computed using the ideal MHD condition. These initial
data satisfy the force-free conditions.

For $a_\ast=0$ we perform a resolution study: the low, medium and high
resolutions cover, $R_{\rm BH}$, the BH apparent horizon ($R_{\rm
  NS}$, the NS minimum) radius by 19, 29, 36 (39, 60, 75) zones,
respectively. The resolutions used for $a_\ast\neq0$ correspond to
the high-resolution $a_\ast=0$ run.  
In all simulations we use 9 levels of refinement with two sets of nested
boxes, differing in size by factors of 2, and each centered onto one
of the orbiting stars. The finest box around the BH (NS) has a side
length $~4.8R_{\rm BH}$ ($~2.4R_{\rm NS}$). We place the outer
boundary at $400M\approx3050(M_{\rm NS}/1.4M_\odot)$km, and impose
reflection symmetry across the orbital plane.


After a transient phase lasting a little over 1 orbit, the B-field
settles into a quasistationary configuration shown in
Fig.~\ref{Fig:Binitial_and_final}. It is evident that for
$a_\ast\neq0$ partial winding of the magnetic field has taken place
due to frame dragging, which is most prominent for $a_\ast=0.75$.


In Fig.~\ref{fig:poyntingfluxdistrib} we show the angular distribution
of the Poynting flux. In all cases, it peaks within a
broad beam of $\sim40^\circ$ in the azimuthal direction, and in the
$a_\ast=0$ and $a_\ast=0.75$ cases within $\sim60^\circ$ from the
orbital plane. This may establish a lighthouse effect as a
characteristic EM signature of BHNS systems prior to merger, if the
variation is not washed out by intervening matter. The distribution of
the flux on a sphere far away from the binary, settles down to an
approximately stationary state in a frame corotating with the binary.

\begin{figure}
\centering
\includegraphics[trim =0.2cm 6.5cm 1.98cm 2.6cm,clip=True,width=0.4\textwidth]{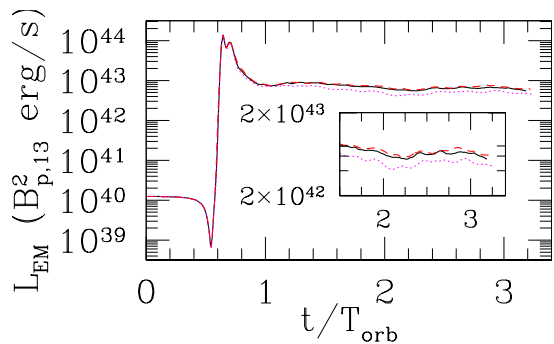}
\caption{Poynting luminosity vs time calculated on a
  sphere of radius $120M=915(M_{\rm NS}/1.4M_\odot)\rm$km for all 3
  cases: $a_\ast = -0.5$ (red) dashed line, $a_\ast = 0$ (black) solid line,
  $a_\ast = 0.75$ (magenta) dotted line.  The inset focuses on the last 1.7
  orbits of evolution.  Here $B_{p,13}=B_{\rm NS,p}/10^{13}\rm G$
  and $T_{\rm orb}$ is the orbital period.  }
\label{Fig:luminosity}
\vspace{-4mm}
\end{figure}

The time evolution of the computed luminosities 
is shown in Fig.~\ref{Fig:luminosity}.  After a transient period
caused by our choice of non-stationary initial magnetic fields, the
luminosities settle to an approximately constant value as expected. We
find that the time-averaged luminosities after the first 1.5 orbits
at the adopted separation are 
\labeq{Lcalc}{
\begin{split}
\langle L_{a_\ast=-0.5}\rangle = &\ 6.6\times 10^{42}\bigg(\frac{B_{\rm NS,p}}{10^{13}\rm G}\bigg)^2\bigg(\frac{M_{\rm NS}}{1.4M_\odot}\bigg)^2 \rm erg/s,\\
\langle L_{a_\ast=0}\rangle = &\ 6.2\times 10^{42}\bigg(\frac{B_{\rm NS,p}}{10^{13}\rm G}\bigg)^2\bigg(\frac{M_{\rm NS}}{1.4M_\odot}\bigg)^2 \rm erg/s,\\
\langle L_{a_\ast=0.75}\rangle = &\ 4.8\times 10^{42}\bigg(\frac{B_{\rm NS,p}}{10^{13}\rm G}\bigg)^2\bigg(\frac{M_{\rm NS}}{1.4M_\odot}\bigg)^2 \rm erg/s,
\end{split}
} 
where $B_{\rm NS,p}$ is the NS polar magnetic field strength measured
by a CTS normal observer, and $M_{\rm NS}$ is the NS rest mass.  As
the B-field does not feed back onto the matter
evolution, the EM luminosity scales exactly as $B^2$.
The characteristic frequency of this EM radiation is of
order the orbital frequency $\sim 200(M_{\rm NS}/1.4M_\odot)^{-1}$Hz
at the adopted separation, and hence smaller than typical
interstellar-medium plasma frequencies $\sim 9$kHz. Thus, this
radiation will be reprocessed before it reaches the observer.

We now compare our results to the approximate UI formula.
The Poynting luminosity of a BHNS UI in the large $q$
limit is given by \cite{McL2011}
%
%
\labeq{LUI2}{
L_{\rm UI} =  \frac{8}{\pi}\bigg(\frac{r_{H}}{2M_{H}}\bigg)^2 v_{\rm rel}^2 \bar B^2_{\rm NS,p}\bigg(\frac{R_{NS}}{r}\bigg)^6q^2M_{\rm NS}^2 
}
where $r_H$ is the horizon radius in units of the BH mass $M_{H}$,
$\bar B_{\rm NS,p}$ is the NS polar magnetic field as measured by
zero-angular-momentum observers (ZAMOs) \cite{MembraneParadigm}, i.e.,
normal observers in a Kerr spacetime in Boyer-Lindquist coordinates,
and $r$ is the binary separation. Here $v_{\rm rel}$ is the azimuthal
velocity of magnetic field lines as measured by ZAMOs, for which the
following relation was proposed \cite{McL2011}: $v_{\rm rel} =
r(\Omega-\Omega_{\rm NS})-\frac{a}{4\sqrt{2}}$,
%
%
where $\Omega$ is the orbital angular frequency, and $\Omega_{\rm NS}$
is the NS spin angular frequency. As our BHNS binaries are
irrotational, we set $\Omega_{\rm NS}=0$. Using the binary parameters
from our simulations and setting $\bar B_{\rm NS,p}\approx B_{\rm
  NS,p}$ in Eq. \eqref{LUI2}, we find
\labeq{LUI2pred}{
\begin{split}
L_{{\rm UI}, a_\ast=0.75} =&\ 0.12\langle L_{a_\ast=0.75}\rangle, \\
L_{{\rm UI}, a_\ast=0} = &\ 0.5\langle L_{a_\ast=0}\rangle, \\
L_{{\rm UI},a_\ast=-0.5} = &\ 0.7\langle L_{a_\ast=-0.5}\rangle.
\end{split}
}
Thus, the UI formula seems to predict well the overall magnitude of
our computed luminosities. However, in contrast to
Eq. \eqref{LUI2pred}, which predicts that $L_{{\rm
    UI},a_\ast=-0.5}/L_{{\rm UI},a_\ast=0}\approx 1.5$ and $L_{{\rm
    UI},a_\ast=-0.5}/L_{{\rm UI},a_\ast=0.75}\approx 7.8$, 
\eqref{Lcalc} shows only a weak dependence of the Poynting
luminosity on the BH spin. This is likely due in part to the spin
dependence being added linearly in the proposed formula for $v_{\rm
  rel}$, and in part to the existence of magnetic dipole emission.

\begin{figure}
\centering
\includegraphics[trim =0.2cm 6.5cm 1.98cm 2.6cm,clip=True,width=0.4\textwidth]{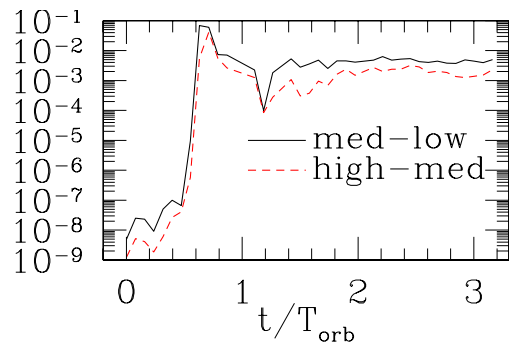}
\caption{Convergence of EM luminosity normalized by the maximum
  luminosity vs.~time. The difference between high and medium
  resolutions is smaller than that between medium and low resolutions,
  indicating that our scheme is convergent.}
\label{Fig:convergence}
\vspace{-4mm}
\end{figure}

In addition to the UI luminosity, another important EM radiation
emission mechanism, that always operates, is that due to the accelerating
MD moment of the NS. 
The approximate MD luminosity is given by
\cite{Ioka:2000yb}
%
\vspace{-0.1cm}
\labeq{}{
\begin{split}
 L_{\rm EM, MD} \approx &\ 2.4\times 10^{41}\bigg(\frac{v}{0.3c}\bigg)^2
              \bigg(\frac{B_{\rm NS,p}}{10^{13}\rm G}\bigg)^2 \\
             &\ \bigg(\frac{M_{\rm NS}}{1.4M_\odot}\bigg)^2\bigg(\frac{r}{6.6R_{\rm NS}}\bigg)^{-6} \rm erg/s, 
\end{split}
} 
where we inserted parameters from our simulations. $L_{\rm EM, MD}$ is
only $\sim 20$ times smaller than what we observe in our simulations,
but is included in our calculated luminosity (see also
\cite{Palenzuela:2013hu}). MD emission dominates when UI ceases
due to corotation (and $a_\ast=0$ for BHNSs) or $\zeta_\phi>1$,
which may be the case for NSNS binaries \cite{DL2012}.

The results of our resolution study for $a_\ast=0$ are shown in
Fig.~\ref{Fig:convergence}, where it is demonstrated that our scheme
is convergent, and that the resulting luminosities in the two highest
resolutions agree to within $\sim 5\%$. Due to numerical resistivity,
the EM energy in the NS interior is conserved after three orbits to within
$10\%,\ 11\%,\ 7\%$ in the $a_\ast=-0.5,\ a_\ast=0,\ a_\ast=0.75$
cases, respectively. Thus, these errors should be taken as the
approximate error bars of our calculations. Our convergence test also
shows that the numerical dissipation decreases toward zero, but the
outgoing radiation converges to a nonzero value, with increasing
resolution. Also, freezing the spacetime and matter evolution, while
evolving the EM fields shows that the outgoing Poynting flux is 4
orders of magnitude smaller than the values in
Eq. \eqref{Lcalc}. Thus, the measured luminosities are not corrupted
by interior energy leaking to the exterior. 
Calculating the ratio of the electric flux to the initial magnetic
flux through a hemisphere of radius $1.5R_{\rm NS}$ centered on the NS
vs time, we find this ratio to be $< 10^{-3}$. 
Furthermore, we performed the low-resolution run of the
nonspinning BH case, setting the initial exterior E-field and $u_i$ to
0. In this case we expect $\int {\bf E}\cdot d{\bf S}=0$. We
calculated the ratio $\int {\bf E}\cdot d{\bf S}/\int |E| dS$, which
quantifies how close to zero $\int {\bf E}\cdot d{\bf S}$ is, and have
found it to be $< 1\%$ at all times. These two results indicate that
little spurious charge is generated in our simulations. Moreover,
calculating the Poynting luminosity in this last run, we find the same
values [within 1\% (0.1\%) following the first (second) orbit] as in
the case where the initial exterior $u_i$ continuously falls off as
$1/r^2$ from its value on the NS surface. Thus, the relaxed solution
we obtain is nearly independent of these initial configurations.

In a future work we plan to extend our simulations to study the
variation of the outgoing Poynting luminosity during the inspiral
phase, and its dependence on different mass ratios. 


\acknowledgments
The authors wish to thank Charles F.~Gammie, Roman Gold, and Yuk Tung Liu for useful discussions. 
We also thank the Illinois Relativity Group's REU team 
[Gregory Colten,
Albert Kim,
Brian Taylor,
and Francis Walsh] for assistance in producing Fig.~\ref{Fig:Binitial_and_final}. These visualizations were
created using the ZIB Amira software package~\cite{Stalling:AmiraVDA-2005}, and we gratefully 
acknowledge the Zuse Institute Berlin for providing us a
license.
This paper was supported in part by NSF Grants AST-1002667, and
PHY-0963136 as well as NASA Grant NNX11AE11G at the
University of Illinois at Urbana-Champaign. This work used the Extreme Science and Engineering Discovery Environment (XSEDE), 
which is supported by NSF grant number OCI-1053575.

\bibliography{paper}

\end{document}